\documentclass[conference]{IEEEtran}
\IEEEoverridecommandlockouts
\usepackage{cite}
\usepackage{amsmath,amssymb,amsfonts}
\usepackage{algorithmic}
\usepackage{graphicx}
\usepackage{textcomp}
\usepackage{xcolor}
\usepackage[acronym]{glossaries}
\makeglossaries
\def\BibTeX{{\rm B\kern-.05em{\sc i\kern-.025em b}\kern-.08em
    T\kern-.1667em\lower.7ex\hbox{E}\kern-.125emX}}
\usepackage{fancyhdr}
\usepackage{svg}
\usepackage{xspace}
\usepackage{booktabs} 
\usepackage{url}
\usepackage{color}
\usepackage{multirow}
\usepackage{enumitem}
\usepackage{array}
\usepackage[nolist,nohyperlinks]{acronym}
\usepackage{footnote}
\usepackage{hyperref}
\hypersetup{colorlinks,allcolors=black}
\acrodef{GI}{gastrointestinal}
\acrodef{AI}{Artificial Intelligence} 
\acrodef{CADx}{Computer Aided Diagnosis} 
\acrodef{ML}{Machine Learning}
\acrodef{DL}{Deep Learning}
\acrodef{CNN}{Convolutional Neural Network}
\acrodef{IOU}[IoU]{Intersection over Union}
\acrodef{CRC}{Colorectal Cancer}
\acrodef{WCE}{Wireless Capsule Endoscopy}  
\acrodef{BoF}{Bag of Feature}  
\acrodef{GIANA}{Gastrointestinal Image ANAlysis} 
\acrodef{FCN}{Fully Convolutional Network}
\acrodef{SGDR}{Stochastic Gradient Descent with Restart}
\acrodef{AUC-ROC}{Area Under Curve - Receiver Operating Characteristic}
\acrodef{ROC}{Receiver Operating Curve}
\acrodef{SGD}{Stochastic Gradient Descent}
\acrodef{MSE}{Mean Square Error}
\acrodef{mIoU}{mean Intersection over Union}
\acrodef{ReLU}{Rectified Linear Unit}
\acrodef{CRC}{Colorectal Cancer}
\acrodef{BN}{Batch Normalization}
\acrodef{ASPP}{Atrous Spatial Pyramid Pooling}
\acrodef{ReLU}{Rectified Linear Unit}
\acrodef{FCN}{Fully Convolutional Network}
\acrodef{mIoU}{mean Intersection over Union}

\begin{document}

\title{DoubleU-Net: A Deep Convolutional Neural Network for Medical Image Segmentation}

\author{\IEEEauthorblockN{Debesh Jha\IEEEauthorrefmark{1}\IEEEauthorrefmark{2}, Michael A. Riegler\IEEEauthorrefmark{1}, 
Dag Johansen\IEEEauthorrefmark{2}, P{\aa}l Halvorsen\IEEEauthorrefmark{1}\IEEEauthorrefmark{3}, H{\aa}vard D. Johansen\IEEEauthorrefmark{2}
}
\vspace{2mm}
\IEEEauthorblockA{\IEEEauthorrefmark{1}SimulaMet, Norway \ \ \ \ \ \
\IEEEauthorrefmark{2}UiT The Arctic University of Norway, Norway \ \ \ \ \ \
\IEEEauthorrefmark{3}Oslo Metropolitan University, Norway \\
Email: {debesh@simula.no}}}

\maketitle

\begin{abstract}
Semantic image segmentation is the process of labeling each pixel of an image with its corresponding class. An encoder-decoder based approach, like U-Net and its variants, is a popular strategy for solving medical image segmentation tasks. To improve the performance of U-Net on various segmentation tasks, we propose a novel architecture called DoubleU-Net, which is a combination of two U-Net architectures stacked on top of each other. The first U-Net uses a pre-trained VGG-19 as the encoder, which has already learned features from ImageNet and can be transferred to another task easily. To capture more semantic information efficiently, we added another U-Net at the bottom. We also adopt \acf{ASPP} to capture contextual information within the network. We have evaluated DoubleU-Net using four medical segmentation datasets, covering various imaging modalities such as colonoscopy, dermoscopy, and microscopy. Experiments on the 2015 MICCAI sub-challenge on automatic polyp detection dataset, the CVC-ClinicDB, the 2018 Data Science Bowl challenge, and the Lesion boundary segmentation datasets demonstrate that the DoubleU-Net outperforms U-Net and the baseline models. Moreover, DoubleU-Net produces more accurate segmentation masks, especially in the case of the CVC-ClinicDB and 2015 MICCAI sub-challenge on automatic polyp detection dataset, which have challenging images such as smaller and flat polyps. These results show the improvement over the existing U-Net model. The encouraging results, produced on various medical image segmentation datasets, show that DoubleU-Net can be used as a strong baseline for both medical image segmentation and cross-dataset evaluation testing to measure the generalizability of \acf{DL} models. 
\end{abstract}

\section{Introduction}
Medical image segmentation is the task of labeling each pixel of an object of interest in medical images. It is often a key task for clinical applications, varying from \ac{CADx} for lesions detection to therapy planning and guidance~\cite{le2015gpssi}. Medical image segmentation helps clinicians focus on a particular area of the disease and extract detailed information for a more accurate diagnosis. The key challenges associated with medical image segmentation are the unavailability of a large number of annotated, lack of high-quality labeled images for training~\cite{jha2020kvasir}, low image quality, lack of a standard segmentation protocol, and a large variations of images among patients~\cite{zhao2013overview}. The quantification of segmentation accuracy and uncertainty is essential to estimate the performance on other applications~\cite{le2015gpssi}. This indicates the requirement for an automatic, generalizable, and efficient semantic image segmentation approach.

\acp{CNN} have shown state-of-the-art performance for automated medical image segmentation~\cite{litjens2017survey}. For semantic segmentation tasks, one of the earlier \acf{DL} architecture trained end-to-end for pixel-wise prediction is a \acf{FCN}. U-Net~\cite{ronneberger2015u} is another popular image segmentation architecture trained end-to-end for pixel-wise prediction. The U-Net architecture consists of two parts, namely, analysis path and synthesis path. In the analysis path, deep features are learned, and in the synthesis path, segmentation is performed based on the learned features.  Additionally, U-Net uses skip connections operation. The skip connection allows propagating dense feature maps from the analysis path to the corresponding layers in the synthesis part. In this way, the spatial information is applied to the deeper layer, which significantly produces a more accurate output segmentation map. Thus, adding more layers to the U-Net will allow the network to learn more representative features leading to better output segmentation masks.

Generalization, i.e., the ability of the model to perform in an independent dataset, and robustness, i.e., the ability of the model to perform on challenging images, are keys  for the development of \ac{AI} system to be used in clinical trials~\cite{ross2020robust}. Therefore, it is essential to design a powerful architecture that is robust and generalizable across different biomedical applications. Pre-trained ImageNet~\cite{deng2009imagenet} models have significantly improved the performance of the \ac{CNN} architectures. One of the examples of such models trained on ImageNet is VGG19~\cite{simonyan2014very}. Inspired by the success of U-Net and its variants for medical image segmentation, we propose an architecture that uses modified U-Net and VGG-19 in the encoder part of the network. Because we use two U-Net architectures in the network, we term the architecture as \textbf{DoubleU-Net}. The main reasons for using the VGG network are: (1)  VGG-19 is a lightweight model as compared to other pre-trained models, (2) the architecture of VGG-19 is similar to U-Net, making it easy to concatenate with U-Net, and (3) it will allow much deeper networks for producing better output segmentation mask. Thus, we aim to improve the overall segmentation performance of the network by enabling this architectural changes. 

The main contributions of this work are: 
\begin{itemize}

\item We propose a novel architecture, DoubleU-Net, for semantic image segmentation. The proposed architecture uses two U-Net architecture in sequence, with two encoders and two decoders. The first encoder used in the network is pre-trained VGG-19~\cite{simonyan2014very}, which is trained on ImageNet~\cite{deng2009imagenet}. Additionally, we use \acf{ASPP}~\cite{chen2017rethinking}. The rest of the architecture is built from scratch. 

\item Experiments on multiple datasets are prerequisites for showing the enhancement of the proposed algorithm over other algorithms. In this respect, we have experimented on four different medical imaging datasets, two different datasets from colonoscopy, one from dermoscopy, and one from microscopy. DoubleU-Net shows better segmentation performance as compared to baseline algorithms on 2015 MICCAI sub-challenge on automatic polyp detection dataset, CVC-ClinicDB dataset, Lesion Boundary Segmentation challenge from ISIC-2018, and 2018 Data Science Bowl challenge dataset. 

\item An extensive evaluation of DoubleU-Net across four dataset shows a significant improvement over U-Net. Therefore, DoubleU-Net can be a new baseline for medical image segmentation task. 

\end{itemize}

The paper is organized into seven sections. Section~\ref{sec:related} provides an overview of the related work in the field of medical image segmentation. In Section~\ref{sec:Doubleunet}, we describe the proposed architecture. Section~\ref{sec:experiment} describes the experiments. Section~\ref{sec:results} presents the results obtained from the experimental evaluation on different datasets. A discussion of the work is provided in Section~\ref{sec:discussion}. Finally, we summarize the paper and discuss future work and limitations in Section~\ref{sec:conclusion}.
\section{Related Work}
\label{sec:related}

Among different \ac{CNN} architectures, an encoder-decoder network like \ac{FCN}~\cite{long2015fully} and its extension U-Net~\cite{ronneberger2015u} have gained significant popularity among semantic segmentation approach for $2$D images. Badrinarayan et al.~\cite{badrinarayanan2017segnet} proposed a deep fully \ac{CNN} for semantic pixel-wise segmentation that has significantly fewer parameters and produces good segmentation maps. Yu et al.~\cite{yu2015multi} proposed a new convolutional network module that particularly targeted dense prediction problems. The proposed module used dilated convolutions for systematically aggregating multi-scale contextual information, and the presented context module improved the accuracy for state-of-the-art semantic image segmentation systems. 

Chen et al.~\cite{chen2017deeplab} proposed DeepLab to solve segmentation problem. Later, DeeplabV3~\cite{chen2017rethinking} significantly improved over their previous DeepLab versions without DenseCRF post-processing. The DeepLabV3 architecture uses a synthesis path that contains the fewer number of convolutional layers that are different from the synthesis path of \ac{FCN} and U-Net. DeepLabV3 uses skip connection between analysis path and synthesis path similar to U-Net architecture. Zhao et al.~\cite{zhao2017pyramid} proposed effective scenes parsing network for complex scene understanding, where global pyramidal features provide an opportunity to capture additional contextual information. Zhang et al.~\cite{zhang2018road} proposed Deep Residual U-Net, which uses residual connections better output segmentation map.
Chen et al.~\cite{chen2018drinet} proposed Dense-Res-Inception Net (DRINET) for medical image segmentation and compared their results with FCN, U-Net, and ResUNet. Ibtehaz et al.~\cite{ibtehaz2020multiresunet} modified U-Net and proposed an improved MultiResUNet architecture for medical image segmentation where they compared their results with U-Net on various medical image segmentation datasets and showed superior accuracy than U-Net. 

Jha et al.~\cite{jha2019resunet++} proposed ResUNet++, which is an enhanced version of standard ResUNet by integrating an additional layer such as squeeze-and-excite block, \ac{ASPP}, and attention block to the network. The proposed architecture uses dice loss as the loss function and produces an improved output segmentation maps as compared to U-Net and ResUNet on the Kvasir-SEG\cite{jha2020kvasir} and CVC-ClinicDB~\cite{bernal2015wm} datasets. Zhou et al.~\cite{zhou2019unet++} proposed UNet++, a neural network architectures for semantic and instance segmentation tasks. They improved the performance of UNet++ by alleviating the unknown network depth, redesigning the skip connections, and devising a pruning scheme to the architecture. 

From the above-related work, we can observe that there has been substantial efforts toward developing deep \ac{CNN} architectures for the segmentation of both natural and medical images. Recently, more works are focused on developing generalizable models, which is why most of the researchers test their algorithms on different datasets~\cite{ibtehaz2020multiresunet,jha2019resunet++,zhou2019unet++}. The accuracy achieved is now is high for both natural imaging~\cite{chen2017deeplab} and medical imaging\cite{ibtehaz2020multiresunet,zhou2019unet++,jha2019resunet++}. 
However, \ac{AI} in medicine is still an emerging field. One of the significant challenges in the medical domain is the lack of test datasets. Moreover, the obtained datasets are often imbalanced. To some extent, we can say that the performance is acceptable in the case of natural images. In the medical imaging, there are many challenging images (for example, flat polyps in colonoscopy), which are usually missed out during colonoscopy examination and can develop into cancer if early detection is not performed. Therefore, there is a need for a more accurate medical image segmentation approach to deal with the challenging images.  Toward addressing this need, we have proposed DoubleU-Net architecture that produces efficient output segmentation masks with the challenging images. 


\section{The DoubleU-Net Architecture}
\label{sec:Doubleunet}
\begin{figure} [!t]
    \centering
    \includegraphics [width= 9cm]{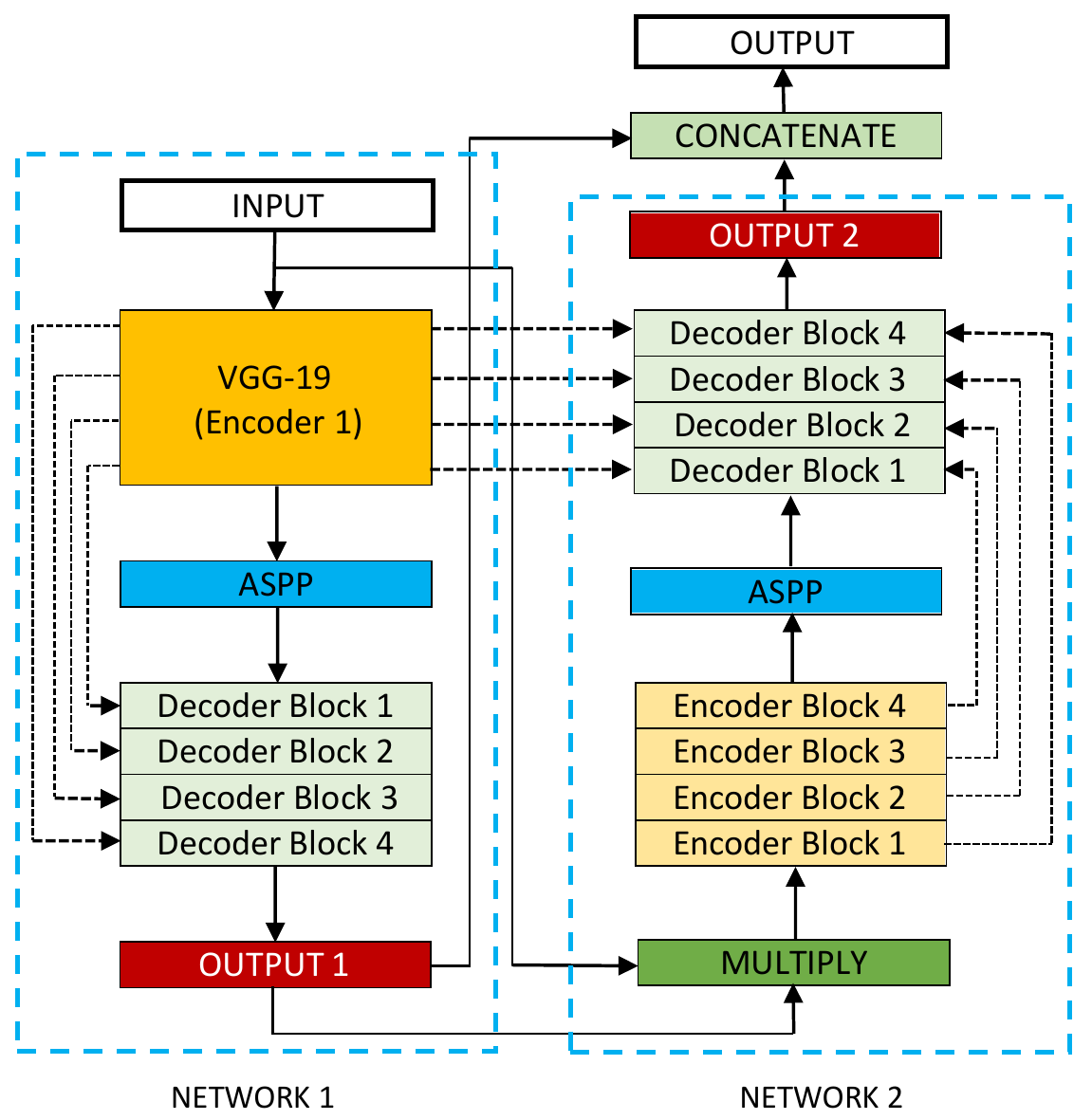}
    \caption{Block diagram of the proposed DoubleU-Net architecture}
    \label{fig:figure1}
\end{figure}

Figure~\ref{fig:figure1} shows an overview of the proposed architecture. As seen from the figure, DoubleU-Net starts with a VGG-19 as encoder sub-network, which is followed by decoder sub-network. What distinguishes DoubleU-Net from U-Net in the first network (NETWORK $1$) is the use of VGG-19 marked in yellow, \ac{ASPP} marked in blue, and decoder block marked in light green. The squeeze-and-excite block~\cite{hu2018squeeze} is used in the encoder of NETWORK $1$ and decoder blocks of NETWORK $1$ and NETWORK $2$. An element-wise multiplication is performed between the output of NETWORK $1$ with the input of the same network. The difference between DoubleU-Net and U-Net in the second network (NETWORK $2$) is only the use of \ac{ASPP} and squeeze-and-excite block. All other components remain the same. 

In the NETWORK $1$, the input image is fed to the modified U-Net, which generates a predicted mask ($Output\-1$). We then multiply the input image and the produced mask ($Output\-1$), which acts as an input for the second modified U-Net that produces another mask ($Output\-2$). Finally, we concatenate both the masks ($Output\-1$ and $Output\-2$) to see the qualitative difference between the intermediate mask ($Output\-1$)  and  final predicted mask ($Output\-2$). 

We assume that the produced output feature map from NETWORK $1$ can still be improved by fetching the input image and its corresponding mask again, and concatenating with $Output 2$ will produce a better segmentation mask than the previous one. This is the main motivation behind using two U-Net architectures in the proposed architecture. The squeeze-and-excite block in the proposed networks reduces the redundant information and passes the most relevant information. \ac{ASPP} has been a popular choice for modern segmentation architecture because it helps to extract high-resolution feature maps that lead to superior performance~\cite{jha2019resunet++}.

\subsection{Encoder Explanation}
The first encoder in DoubleU-Net ($encoder\-1$) uses pre-trained VGG-19, whereas the second encoder ($encoder\-2$), is built from scratch. Each encoder tries to encode the information contained in the input image. Each encoder block in the $encoder\-2$ performs two $3\times3$ convolution operation, each followed by a batch normalization. The batch normalization reduces the internal co-variant shift and also regularizes the model. A \ac{ReLU} activation function is applied, which introduces non-linearity into the model. This is followed by a squeeze-and- excitation block, which enhances the quality of the feature maps. After that, max-pooling is performed with a $2\times2$ window and stride $2$ to reduce the spatial dimension of the feature maps.   
\vspace{-0.2cm}
\subsection{Decoder Explanation}
As shown in Figure~\ref{fig:figure1}, we use two decoders in the entire network, with small modifications on the decoder as compared with that of the original U-Net. Each block in the decoder performs a $2\times2$ bi-linear up-sampling on the input feature, which doubles the dimension of the input feature maps. Now, we concatenate the appropriate skip connections feature maps from the encoder to the output feature maps. In the first decoder, we only use skip connection from the first encoder, but in the second decoder, we use skip connection from both the encoders, which maintains the spatial resolution and enhance the quality of the output feature maps. After concatenation, we again perform two $3\times3$ convolution operation, each of which is followed by batch normalization and then by a \ac{ReLU} activation function. After that, we use a squeeze and excitation block. At last, we apply a convolution layer with a sigmoid activation function, which is used to generate the mask for the corresponding modified U-Net.

\section{Experiments}
\label{sec:experiment}
In this section, we present datasets, evaluation metrics, experiment setup and configuration, and data augmentation techniques used in all the experiments to validate the proposed framework.  

\subsection{Datasets}
To evaluate the effectiveness of the DoubleU-Net, we have used four publicly available datasets from medical domain.
\begin{itemize}

\item The 2015 MICCAI sub-challenge on automatic polyp detection~\cite{bernal2017comparative} used the CVC-ClinicDB~\cite{bernal2015wm} for training and ETIS-Larib~\cite{silva2014toward} for testing in the case of polyp detection task. The 2015 MICCAI sub-challenge on automatic polyp detection dataset is the first dataset used in our study. 

\item Similarly, CVC-ClinicDB has been a common choice for polyp segmentation.  Therefore, we use this dataset for comparison. 

\item The third dataset used in our experiment is from the ISIC-2018 challenge, namely, Lesion Boundary Segmentation dataset~\cite{codella2018skin,tschandl2018ham10000}. The dataset contains skin lesions and their corresponding annotations. 

\item The fourth dataset used in this study is nuclei segmentation, from the 2018 Data Science Bowl challenge\footnote{\url{https://www.kaggle.com/c/data-science-bowl-2018}}. This dataset is publicly available at Broad Bioimage Benchmark Collection\footnote{\url{https://data.broadinstitute.org/bbbc/BBBC038/}}.

\end{itemize}
More information about the datasets are presented in Table~\ref{table:dataset}. All of the datasets are clinically relevant during diagnosis, and therefore, their segmentation can be crucial for patient outcome.

\begin{table} [!t]
 \caption{Summary of biomedical segmentation dataset used in our experiments}
    \label{table:dataset}
  \def\arraystretch{1.1}
    \setlength\tabcolsep{5pt}
    \par\bigskip
    \centering
        \resizebox{\columnwidth}{!}{%
          \begin{tabular}{ l c c c c} 
                \toprule
                Dataset & \shortstack{No. of\\ Images} & Input size & Application\\ 
              \bottomrule
                2015 MICCAI sub-challenge on automatic polyp detection dataset  & 808 & $384\times 288$ & Colonoscopy\\ 
                 CVC-ClinicDB & 612 & $384\times 288$ & Colonoscopy \\ 
                 Lesion Boundary Segmentation challenge & 2594 &Variable & Dermoscopy\\ 
                 2018 Data Science Bowl Challenge & $670$ &$256\times256$ & Nuclei\\ 
                 \bottomrule
\end{tabular}}
\end{table}

\subsection{Evaluation metrics}
DoubleU-Net is evaluated on the basis of S{\o}rensen–dice coefficient (DSC), \acf{mIoU}, Precision, and Recall. We evaluate all of these metrics for all four datasets. However, we compare and emphasize more on the official evaluation metrics that were used in the challenge. For example, the official evaluation metrics for the Lesion Boundary Segmentation challenge is \ac{mIoU}. 

\subsection{Experiment setup and configuration}
All models are implemented using Keras framework~\cite{chollet2015keras} with Tensorflow~$2.1.0$~\cite{abadi2016tensorflow} as backend. The implementation can be found at our GitHub repository\footnote{\url{https://github.com/DebeshJha/2020-CBMS-DoubleU-Net}}. We ran our experiments on a Volta 100 GPU and an Nvidia DGX-2 AI system. In all of the datasets, we used 80\% of dataset for training, 10\% for validation, and 10\% for testing. During training, we used the original image size for the smaller dataset, such as CVC-ClinicDB and Nuclei segmentation dataset, and resized the images to $384\times512$ for the Lesion Boundary segmentation challenge dataset to balance between training time and complexity. The size of ETIS-Larib was adjusted similarly to that of CVC-ClinicDB. We use binary cross-entropy as the loss function for all the networks and the Nadam optimizer with its default parameters. For the lesion boundary segmentation dataset and the Nuclei segmentation dataset, where dice loss and Adam optimizer performed slightly higher, the batch size is set to $16$ and the learning rate to $1\mathrm{e}{-5}$. All models are trained for $300$ epochs. Early stopping and ReduceLROnPlateau is also used.  

\subsection{Data augmentation techniques}
Medical datasets are challenging to obtain and annotate~\cite{jha2020kvasir}. Most existing datasets have only a few samples, which makes  training \ac{DL} models on these datasets challenging. One potential solution to the challenge of data insufficiency, is to use data augmentation techniques that increase the number of samples during training. For this, we first split the dataset into training, validation, and testing sets. We then apply different data augmentation methods to each set, including center crop, random rotation, transpose, elastic transform, etc. More details about the augmentation techniques we used can be found in our GitHub repository. A single image was converted into $25$ different images; thus, in total, $26$ images including the original image.  The same augmentation techniques were applied to all four datasets.

\begin{table}[t]
 \caption{Experimental results using the 2015 MICCAI sub-challenge on automatic polyp detection dataset}
    \label{table:result1}
   \def\arraystretch{1.1}
    \setlength\tabcolsep{5pt}
    \par\bigskip
    \centering
        \resizebox{\columnwidth}{!}{%
           \begin{tabular}{ l c c c c} 
                \toprule
                Method & DSC & \ac{mIoU} & Recall & Precision\\ 
              \bottomrule
                 FCN-VGG~\cite{brandao2017fully} &0.7023 &0.5420  &- &- \\ 
                 Mask R-CNN with Resnet101~\cite{qadir2019polyp} &0.7042 &0.6124  &- &- \\ 
                 U-Net & 0.2920 & 0.1759 &0.5930 &0.2021\\ 
                 DoubleU-Net & \textbf{0.7649} &\textbf{0.6255} &\textbf{0.7156} &\textbf{0.8007} \\ 
                 \bottomrule
\end{tabular}}
\end{table}						


\begin{figure} [t]
    \centering
    \includegraphics [width=9cm ]{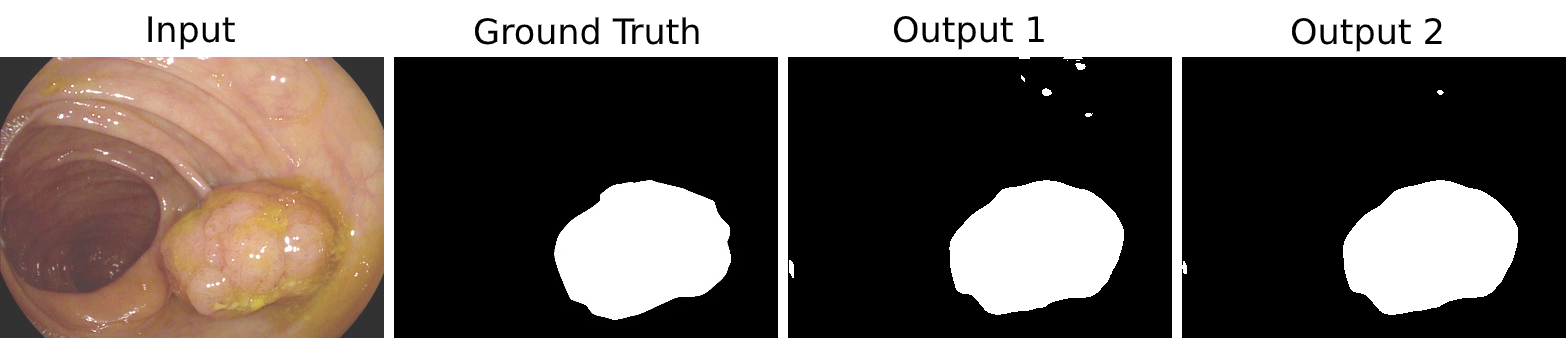}\vspace{0.5mm}\\
    \includegraphics [width=9cm ]{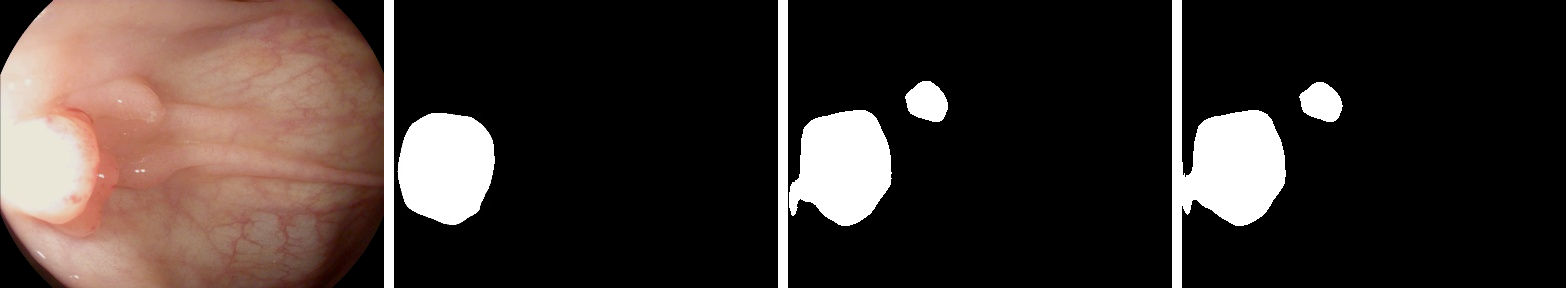}\vspace{0.5mm}\\
    \includegraphics [width=9cm ]{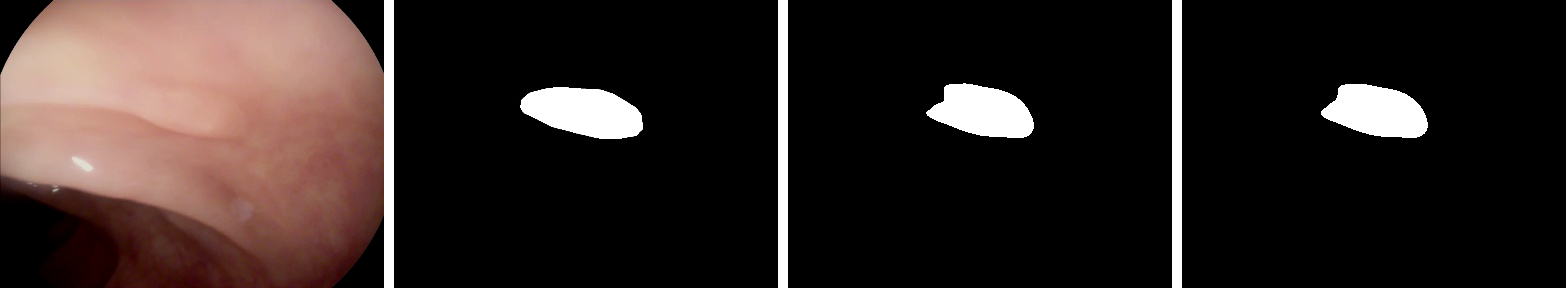}\vspace{0.5mm}\\
    \includegraphics [width=9cm]{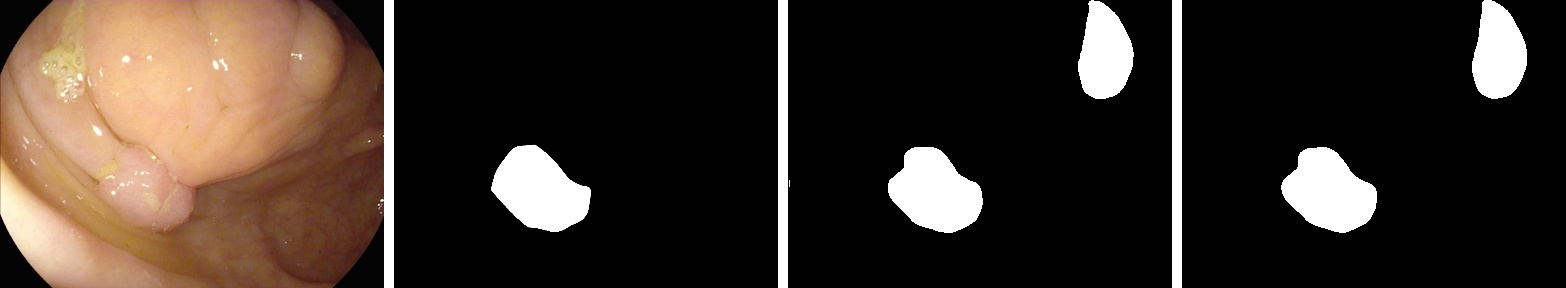}\vspace{0.5mm}\\
    \caption{Qualitative result of DoubleU-Net on large, medium, and flat polyps from 2015 MICCAI sub-challenge on automatic polyp detection dataset}
  \label{fig:figure2}
\end{figure}

\begin{table} [t]
 \caption{Result comparison on CVC-ClinicDB}
    \label{table:result2}
   \def\arraystretch{1.1}
    \setlength\tabcolsep{5pt}
    \par\bigskip
    \centering
        \resizebox{\columnwidth}{!}{%
           \begin{tabular}{ l c c c c} 
                \toprule
                Method & DSC & mIoU & Recall & Precision\\ 
              \bottomrule
                Fully Convoutional Network~\cite{li2017colorectal} & - &- &0.7732 & 0.8999\\ 
                CNN~\cite{nguyen2018colorectal} & (0.62-0.87) &- &- & -\\
                SegNet~\cite{wang2018development} & - & - & \textbf{0.8824} & -\\ 
                {Multi-scale patch-based CNN}~\cite{banik2020multi} & 0.8130 &- &0.7860 & 0.8090\\
                MultiResUNet with data augmentation~\cite{ibtehaz2020multiresunet}&  - & 0.8497 & - & -\\
                Conditional generative adversarial network\cite{poomeshwaran2019polyp} & 0.8848 & 0.8127 & - & -\\ 
                U-Net &  0.8781 & 0.7881 & 0.7865 & 0.9329\\
                DoubleU-Net &\textbf{0.9239} &\textbf{0.8611 }&0.8457 &\textbf{0.9592}\\ 
                 \bottomrule
\end{tabular}}
\end{table}

\begin{figure} [t]
    \centering
    \includegraphics [width=9cm ]{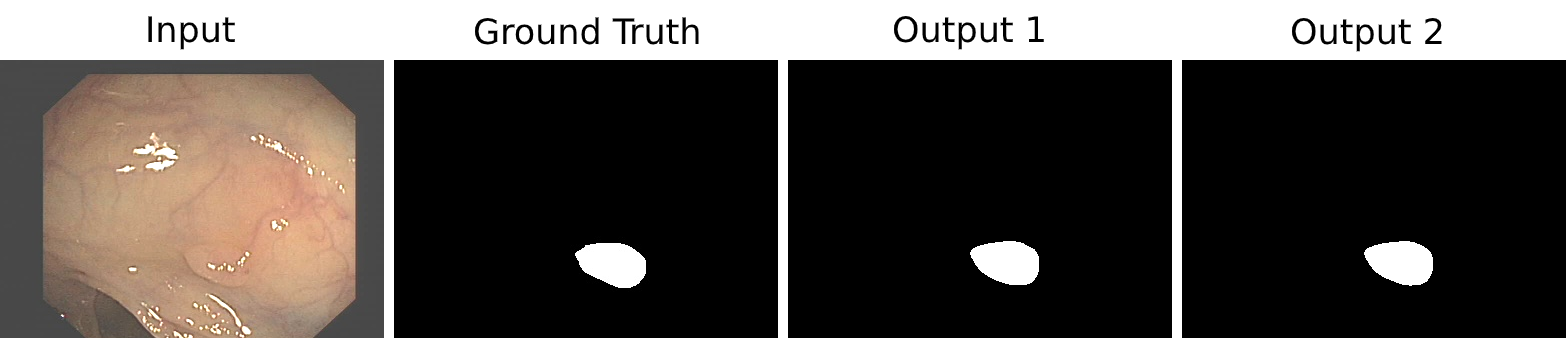}\vspace{0.5mm}\\
    \includegraphics [width=9cm ]{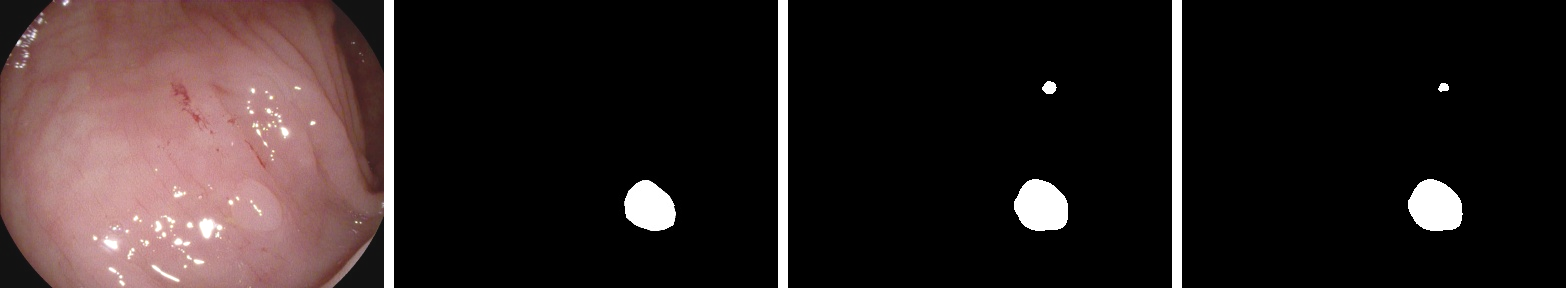}\vspace{0.5mm}\\
    \includegraphics [width=9cm ]{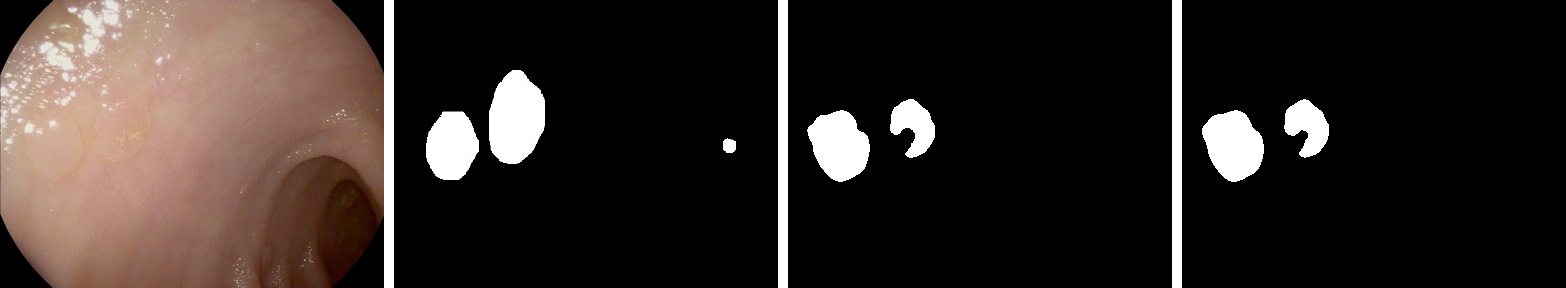}\vspace{0.5mm}\\
    \includegraphics [width=9cm]{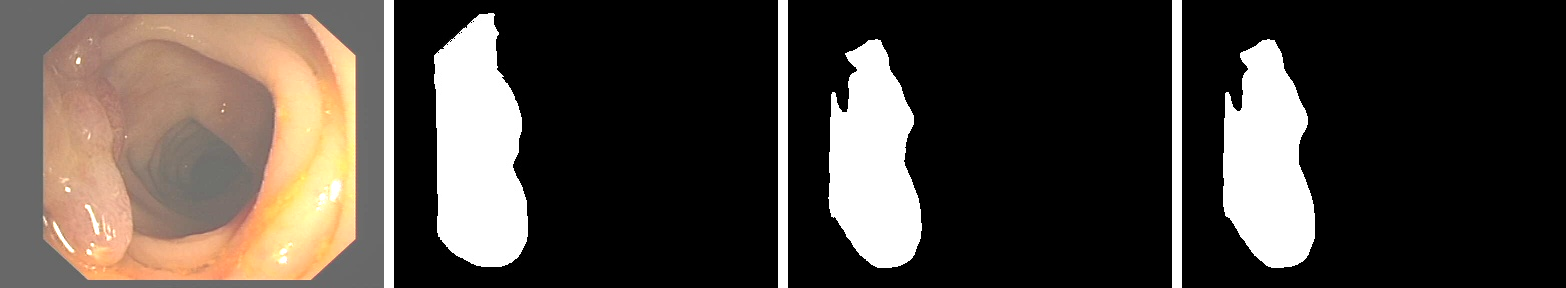}\vspace{0.5mm}\\
    \caption{Qualitative result of DoubleU-Net on challenging images from CVC-ClinicDB}
  \label{fig:figure3}
\end{figure}


\begin{table} [b]
 \caption{Result on Lesion boundary segmentation dataset from ISIC-2018}
    \label{table:results3}
   \def\arraystretch{1.1}
    \setlength\tabcolsep{5pt}
    \par\bigskip
    \centering
        \resizebox{\columnwidth}{!}{%
           \begin{tabular}{ l c c c c} 
                \toprule
                Method & DSC & \ac{mIoU} & Recall & Precision\\ 
              \bottomrule
                 U-Net\cite{ibtehaz2020multiresunet}  & - &$0.7642 \pm 0.4518$ &- &- \\ 
                 Multi-ResUNet\cite{ibtehaz2020multiresunet} &- &$0.8029 \pm 0.3717$&- &-\\ 
                 DoubleU-Net &\textbf{0.8962}  &\textbf{0.8212} & \textbf{0.8780} &\textbf{0.9459}\\ 
                 \bottomrule
\end{tabular}}
\end{table}	

\begin{figure} [t]
    \centering
    \includegraphics [width=9cm]{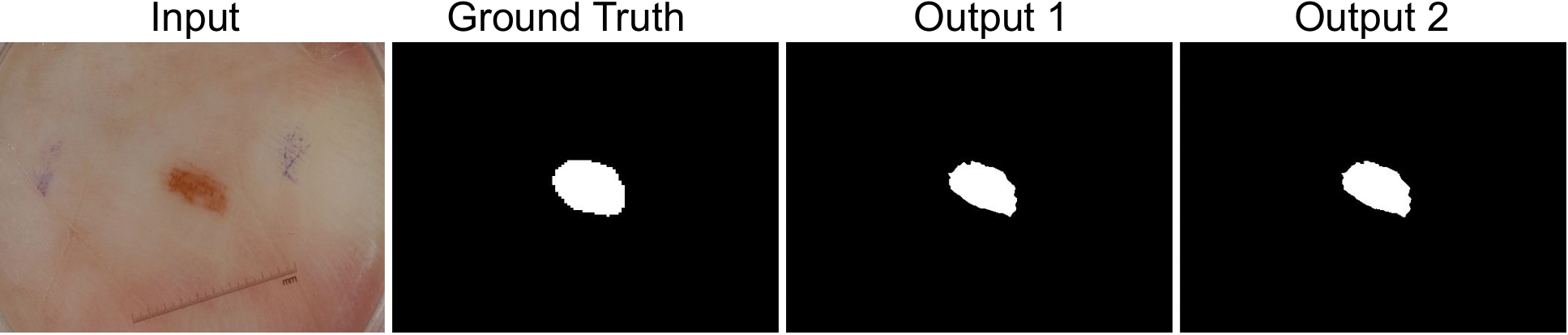}\vspace{0.5mm}\\
    \includegraphics [width=9cm]{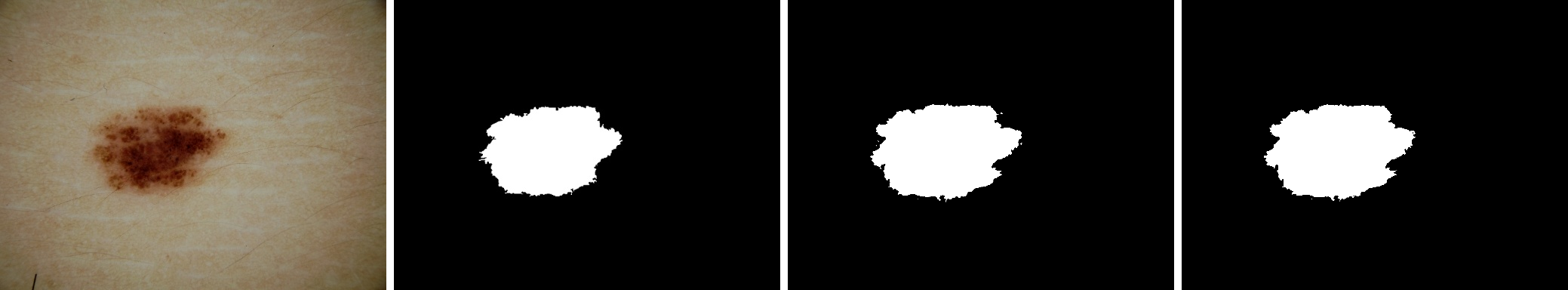}\vspace{0.5mm}\\
    \includegraphics [width=9cm]{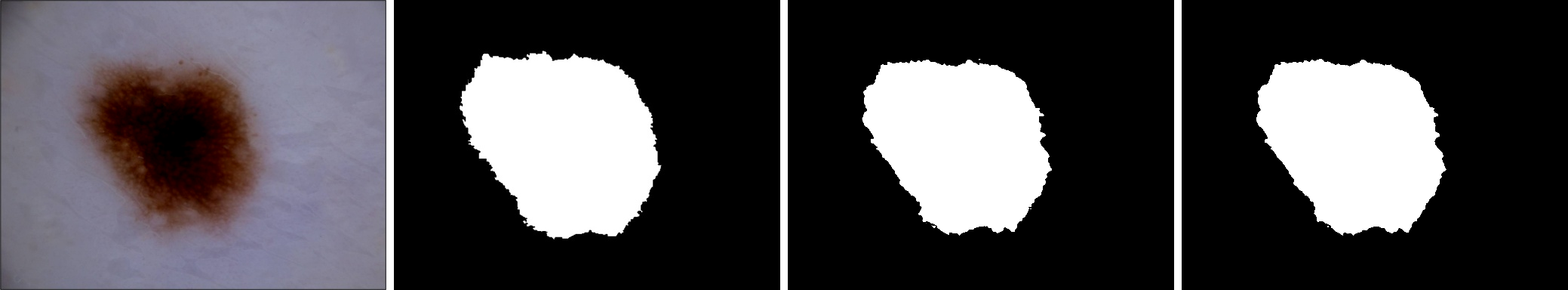}\vspace{0.5mm}\\
    \includegraphics [width=9cm]{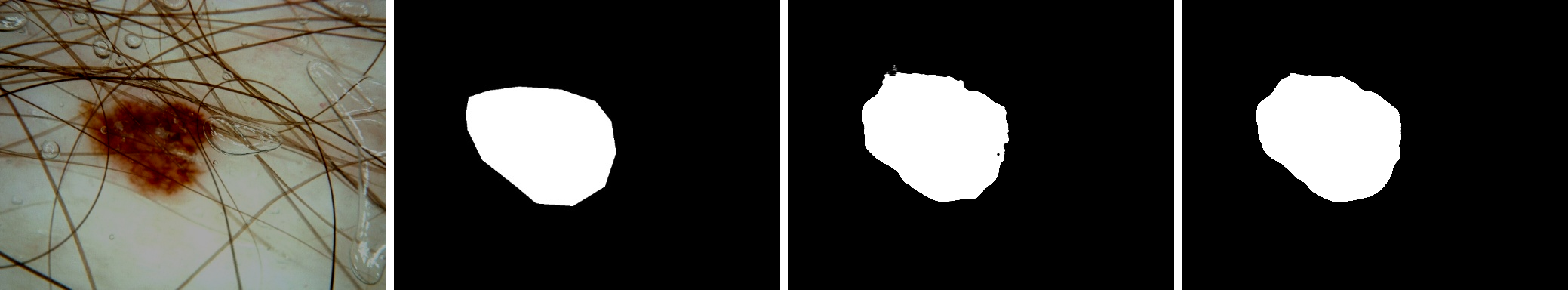}\vspace{0.5mm}\\
    \caption{Qualitative result of DoubleU-Net on small, medium and large size skin lesions from Lesion Boundary segmentation challenge}
  \label{fig:figure4}
\end{figure}

\section{Results}
\label{sec:results}
In this section, we present the results and compare them with the baselines on the respective datasets. U-Net is still considered as the baseline for various medical image segmentation tasks. Therefore, we compare the proposed model with U-Net by using the same data augmentation techniques as described above to demonstrate its effectiveness.  We also report the results on four datasets and show the qualitative results to prove the usefulness of DoubleU-Net. In all of the figures demonstrating the qualitative results, the sequence of input, ground truth, $Output\-1$, and $Output\-2$ are followed, where $Output\-1$ and $Output\-2$ are the intermediate and final output respectively. 

\subsection{Comparison on 2015 MICCAI sub-challenge on automatic polyp detection dataset}
Our quantitative results on the 2015 MICCAI sub-challenge on automatic polyp detection dataset are summarized in Table~\ref{table:result1}.  The experimental results shows that DoubleU-Net achieved a DSC of $0.7649$ and a \ac{mIoU} of $0.6255$. From  Table~\ref{table:result1}, we can see that DoubleU-Net outperforms the baseline~\cite{qadir2019polyp} by $6.07\%$ in terms of DSC and $1.31\%$ in \ac{mIoU}. From the above table, we can also observe that the model that uses a pre-trained ImageNet network (for instance, Resnet101 or VGG-16) as a backbone achieves a higher score on cross-dataset evaluation as compared to that of training a network from scratch (see Table~\ref{table:result1}). The visual results of the proposed model can be seen in Figure \ref{fig:figure2}. From the visual analysis, we can observe that the segmentation mask produced by $Output\-2$ is better than that of $Output\-1$. This also justifies the significance of the proposed model over U-Net. 

\subsection{Comparison on CVC-ClinicDB}
DoubleU-Net is compared with U-Net and the recent works that used the same dataset for evaluation. Table~\ref{table:result2} shows the results on CVC-ClinicDB dataset. The evaluation results shows that DoubleU-Net achieve a DSC of $0.9239$ which is $3.91\%$ higher than~\cite{poomeshwaran2019polyp} and \ac{mIoU} of $0.8611$, which is $1.14\%$ higher than~\cite{ibtehaz2020multiresunet}. A careful visual analysis of the result shows that DoubleU-Net produces better segmentation masks as compared to the intermediate network. The model performs reasonably well on the challenging images such as flat and small polyps, which are usually missed-out during colonoscopy examinations (see Figure~\ref{fig:figure3}).

\subsection{Comparison on Lesion Boundary segmentation challenge dataset}
The official evaluation metric for the challenge was \ac{mIoU}. DoubleU-Net achieve a DSC of $0.8962$ and \ac{mIoU} of $0.8212$ on this challenge dataset. From the quantitative results comparison (see Table~\ref{table:results3}), we can see that the DoubleU-Net outperforms U-Net~\cite{ibtehaz2020multiresunet} by an approximate margin of $5.7\%$, and Multi-ResUNet~\cite{ibtehaz2020multiresunet} by an approximate margin of $1.83\%$ in terms of \ac{mIoU} on Lesion boundary segmentation challenge dataset from ISIC-2018. Figure~\ref{fig:figure4} shows the qualitative results. From the figure, we can see that both intermediate output and the final output produced by the network perform well on all types of lesions ranging from small to medium to large lesions. However, a careful analysis shows that the final output produced by the network is better than the intermediate one.  

\subsection{Comparison on 2018 Data Science Bowl challenge dataset}
\begin{figure} [t]
    \centering
    \includegraphics [width=9cm, height =1.7cm ]{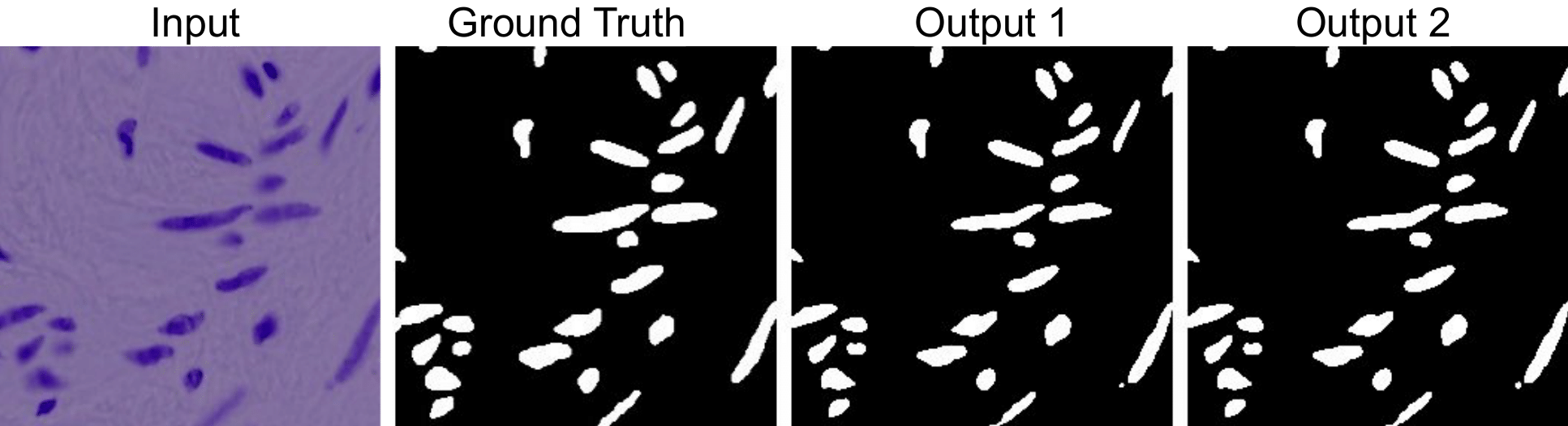}\vspace{0.5mm}\\
    \includegraphics [width=9cm, height =1.7cm ]{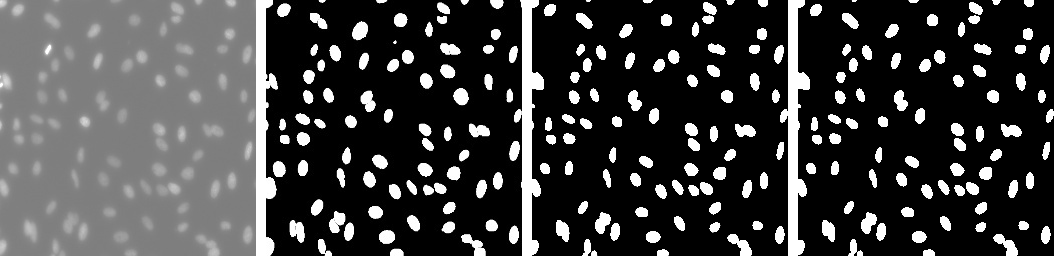}\vspace{0.5mm}\\
    \includegraphics [width=9cm, height =1.7cm ]{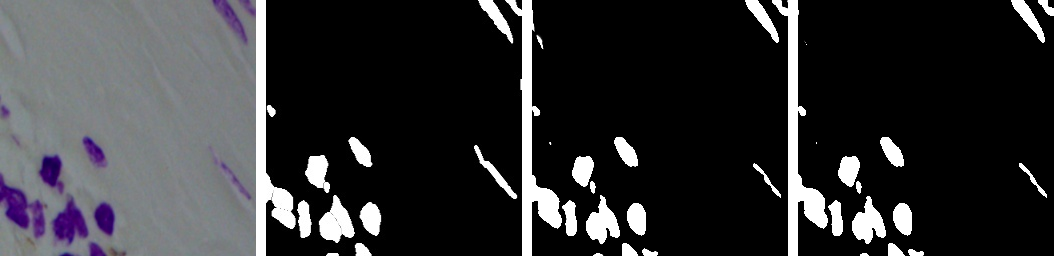}\vspace{0.5mm}\\
    \includegraphics [width=9cm, height =1.7cm ]{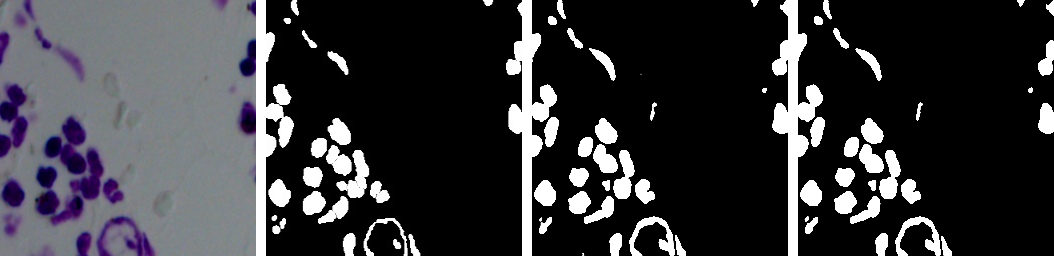}\vspace{0.5mm}\\
    \caption{Qualitative result of DoubleU-Net on Nuclei images from 2018 Data Science Bowl challenge dataset}
  \label{fig:figure5}
\end{figure}

Table~\ref{table:results4} and Figure~\ref{fig:figure5} presents the quantitative and qualitative results on 2018 Data Science Bowl challenge dataset. We have compared our work with U-Net++~\cite{zhou2019unet++}. Our method produced a DSC of $0.9133$, which is $1.59\%$ higher than the method proposed by Zhou et. al~\cite{zhou2019unet++}, and comparable \ac{mIoU} with U-Net and UNet++ that uses Resnet101 as the backbone model. UNet++ has been used as a strong baseline for result comparison over various image segmentation tasks. Therefore, the DoubleU-Net set a new baseline for semantic image segmentation task.

\begin{table} [t]
 \caption{Result on Nuclei segmentation from 2018 Data Science Bowl challenge}
 
    \label{table:results4}
  \def\arraystretch{1.1}
    \setlength\tabcolsep{5pt}
    \par\bigskip
    \centering
        \resizebox{\columnwidth}{!}{%
          \begin{tabular}{ l c c c c c} 
                \toprule
                Method & Pre-trained network & DSC & \ac{mIoU} & Recall & Precision\\ 
              \bottomrule
                  U-Net~\cite{zhou2019unet++} &Resnet101 & $0.7573$ &$0.9103$ &- &- \\ 
                 UNet++~\cite{zhou2019unet++} &Resnet101 & $0.8974$ & $\textbf{0.9255}$ &- &- \\ 
                 DoubleU-Net &VGG-19& $\textbf{0.9133}$& $0.8407$ &$\textbf{0.6407}$ &\textbf{0.9496}\\ 
                 \bottomrule
\end{tabular}}
\end{table}

\begin{table} [t]
 \caption{Relative improvement of DoubleU-Net on U-Net}
    \label{table:resultscomparison}
  \def\arraystretch{1.1}
    \setlength\tabcolsep{5pt}
    \par\bigskip
    \centering
        \resizebox{\columnwidth}{!}{%
          \begin{tabular}{ l c c c c} 
                \toprule
                Modality & {\shortstack{U-Net\\(DSC)}} &{\shortstack{DoubleU-Net\\(DSC)}} & {\shortstack{Overall\\Improvement}}\\ 
              \bottomrule
                Colonoscopy (MICCAI 2015) &$0.2920$ &$0.7649$ & $0.4729$ \\ 
                Colonoscopy (CVC-ClinicDB) &$0.8781$ &$0.9239$ & $0.0458$\\ 
                Dermoscopy (ISIC-2018) &$-$ &$0.8962$ & $-$\\ 
                
                Microscopy (2018 Data Science Bowl ) & $0.7573$ &$0.9133$ &$0.1560$ \\ 
                 \bottomrule
\end{tabular}}
\end{table}

\section{Discussion}
\label{sec:discussion}
Table~\ref{table:resultscomparison} shows the DSC comparison of U-Net and DoubleU-Net. From the above table, we can see that DoubleU-Net performs reasonably well as compared to U-Net for all the presented datasets. For the CVC-ClinicDB dataset, the performance of U-Net is competitive. However, for 2015 MICCAI sub-challenge on automatic polyp detection dataset and the 2018 Data Science Bowl, DoubleU-Net has a significant DSC improvement of $0.4729\%$ and $15.60\%$ respectively. Additionally, the 2015 MICCAI sub-challenge on automatic polyp detection dataset provides us the opportunity to study the cross-data generalizability,  which is critical in the medical domain~\cite{thambawita2020extensive}.   The generalization test showed that DoubleU-Net outperforms its competitors (see Table \ref{table:result1}). From the Table, we observe that the model trained on pre-trained ImageNet~\cite{deng2009imagenet} performs much better on the cross-dataset test than that of the model trained from scratch. We have trained U-Net on the CVC-ClinicDB dataset, which is competitive with DoubleU-Net when tested on the same dataset (see Table~\ref{table:result2}). The same model was used to test against the ETIS-Larib dataset. However, the performance of the U-Net was poor as compared to that of DoubleU-Net (see Table~\ref{table:result1}).  This fact suggests that DoubleU-Net is more generalizable and can be used for the cross-dataset test across the different domains. 

From the qualitative results, we can see that DoubleU-Net is capable of producing better segmentation mask even for the challenging images. This can be observed from Figure~\ref{fig:figure2} and Figure~\ref{fig:figure3}. Moreover, Figure~\ref{fig:figure4} and Figure~\ref{fig:figure5} show that the model produces high-quality segmentation masks for Lesion Boundary Segmentation challenge dataset and 2018 Data Science Bowl challenge dataset. The overall qualitative result shows that the model performs well for different multi-organ and multi-centered medical image segmentation datasets. Thus, the above results suggest that the robustness of the proposed model. 

From the above experiments, we observed that the transfer learning from a pre-trained ImageNet network significantly improves the results on every dataset, which tries to compensate for the lack of enough training data. The qualitative and quantitative results suggest using DoubleU-Net as a baseline for result comparisons over four medical image segmentation datasets.

\section{Conclusion}
\label{sec:conclusion}
In this paper, we have proposed a novel \ac{CNN} architecture called DoubleU-Net. The DoubleU-Net has five main components, namely two U-Net networks, VGG-19, a squeeze-and-excite block and \ac{ASPP}. The performance of DoubleU-Net is significantly better when compared with the baselines and U-Net on all four datasets.

Moreover, the proposed architecture is flexible, and that makes it possible to integrate other \ac{CNN} blocks into DoubleU-Net. We believe that the segmentation results can be improved by further integrating different \ac{CNN} blocks and by the use of post-processing techniques such as conditional random field and Otsu threshold. 

In the future, we plan to research building one model for different medical image segmentation tasks and focus on simplifying the architecture while retaining its ability to produce high segmentation masks. A limitation of the DoubleU-Net is that it uses more parameters as compared to U-Net, which leads to an increase in the training time. In the future, the research should focus more on designing simplified architectures with fewer parameters while maintaining its ability. 
\section*{Acknowledgement}
This work is funded in part by Research Council of Norway project number 263248 (Privaton). The computations in this paper were performed on equipment provided by the Experimental Infrastructure for Exploration of Exascale Computing (eX3), which is financially supported by the Research Council of Norway under contract 270053.

\bibliographystyle{IEEEtran}
\bibliography{references} 

\end{document}